%% file: wellposed_may30.tex
\input macros

\def\sec{}
\centerline{\bf Problems which are well-posed in a generalized sense}
\centerline{\bf with applications to the Einstein equations}
\medskip\centerline{by}
\centerline{H.-O. Kreiss${}^{1,2}$ and J. Winicour${}^{2,3}$ }
\medskip

${}^{1}$NADA, Royal Institute of Technology, 10044 Stockholm, Sweden

${}^{2}$Albert Einstein Institute, Max Planck Gesellschaft,
Am M\"uhlenberg 1, D-14476 Golm, Germany

${}^{3}$Department of Physics and Astronomy,
University of Pittsburgh, Pittsburgh, Pennsylvania 15260
\bigskip

\centerline{\bf Summary}
\medskip
In the harmonic description of general relativity, the principle part of
Einstein equations reduces to a constrained system of 10  curved space wave
equations for the components of the space-time metric. We use the
pseudo-differential theory  of systems which are strongly well-posed in the
generalized sense to establish the well-posedness of constraint preserving
boundary conditions for this system when treated in second order differential
form. The boundary conditions are of a generalized Sommerfeld type that is
benevolent for numerical calculation.

\bigskip
\centerline{\bf 1. Introduction}
\medskip
Consider a first order symmetric system of partial differential
equations
$$
  u_t=Au_x+Bu_y+F 
$$
on the half-space $ x\ge 0,~-\infty<y<\infty,~t\ge 0.$ Here $u$ is a
vector valued function with $n$ components and $A=A^*,~B=B^*$ are
symmetric matrices which depend smoothly on $x,y$ and $t.$ Also,
$A$ is not singular at $x=0.$ At $t=0,$ we give the initial condition
$$ u(x,y,0)=f(x)$$
and, at $x=0,$ boundary conditions
$$ Bu=g $$
which are strictly maximally dissipative.
In this case one can use integration by parts to derive an energy
estimate. The result can easily be extended to prove local existence
of solutions of quasilinear systems because integration by parts
allows us to estimate the derivatives.

This result depends heavily on integration by parts. If the boundary conditions
are not maximally dissipative or the system is not symmetric hyperbolic, new
techniques are needed. There is a rather comprehensive theory based on the
principle of frozen coefficients, Fourier and Laplace transform and the theory
of pseudo-differential operators which give necessary and sufficient conditions
for well-posedness in the generalized sense [1],[2],[3],[4],[5, Chap.8],[6,
Chap 10],[7].  This theory can also be extended to second order systems. In the
context of pseudo-differential operators one can -- in the same way as for
ordinary differential equations -- write a second order system as a first order
system [8].

To make the results of the above theory more precise and how to apply it we
have included an Appendix, which should make it easier to read the paper.

In this paper we shall demonstrate that flexibility of the permissible boundary
conditions can be applied to solve the constraint problem for the harmonic
Einstein equations. The pseudo-differential theory is applied here to the
second order harmonic formulation, which was used to establish the first
well-posed Cauchy problem for Einstein's equations [9].

The importance of a well-posed constraint preserving initial-boundary value
problem (IBVP) to the simulation of Einstein equations has been
recognized in numerous recent works. The pseudo-differential treatment of the
IBVP presented here is most similar to treatments of first order formulations
by Stewart [10], Reula and Sarbach [11] and Sarbach and Tiglio [12]. 
 A well-posed IBVP for the nonlinear harmonic Einstein
equations has been formulated for a combination of homogeneous Neumann and
Dirichlet boundary conditions (or boundary data linearized off these
homogeneous conditions) [13]. Here we  consider strictly dissipative,
Sommerfeld-type boundary conditions for these harmonic equations,
which have proved to be more robust in numerical tests [14,15]. The only general
treatment of the nonlinear case has been given by Friedrich and Nagy, based
upon a quite different first order formulation of the Einstein equations [16]

In the next section we will explain our theory for the wave equation with
boundary conditions which cannot be treated by integration by parts.
We apply the results in the third section to the constraint problem
of the linearized Einstein equation.

In the last section we will give a more physical interpretation of our
technique and explain how it applies to the full Einstein theory.
\vfill\eject
\bigskip
\centerline{\bf 2. Well-posed problems}
\medskip
Consider the half-plane problem for the wave equation
$$
v_{tt}=v_{xx}+v_{yy}+F,\quad 0\le x <\infty,~-\infty <y<\infty,~ t\ge 0,
\eqno(1)
$$
with smooth bounded initial data
$$
v(x,y,0)=f_1(x,y),\quad v_t(x,y,0)=f_2(x,y),
\eqno(2) $$
and boundary condition
$$
v_t=\alpha v_x+\beta v_y+g ,\quad x=0,~-\infty < y < \infty,~t\ge 0.
\eqno(3) $$
Here $\alpha,\beta$ are real constants.
We assume always that $\alpha>0.$ Also, all data are $C^\infty-$smooth
compatible functions with compact support.

The usual concept of well-posedness is based on the existence of an
energy estimate which is often derived by integration by parts. Let
$F\equiv g\equiv 0$ and
$$ \|v(\cdot,\cdot,t)\|^2=\int_0^\infty\!\int_{-\infty}^\infty
|v(x,y,t)|^2 dx\,dy
$$
denote the usual $L_2$-norm. Integration by parts gives us , for the
energy $E,$
$$
\partt E=: \partt (\|v_t\|^2+ \|v_x\|^2+ \|v_y\|^2)=
-2\int_{-\infty}^\infty  v_t(0,y,t) v_x(0,y,t) dy
$$
If in (3) $\beta \ne 0,$ then
there is no obvious way to estimate the boundary flux in terms of $E.$
Of course, if $\beta=0$ and $\alpha >0,$ there is an energy estimate.
To be able to discuss the general case we will define well-posedness
 as in the Appendix.

We start with a very simple observation:
\proclaim Lemma 1. The problem is not well-posed if we can find a solution
of the homogeneous equation (1) which satisfies the homogeneous
boundary condition (3) and which is of the form
$$ v(x,y,t)=e^{st+i\om y} \varphi(x),\quad |\varphi|_\infty <\infty. \eqno(4) $$
Here $\varphi(x)$ is a smooth bounded function and
$\om$ is real and $s=i\xi +\eta$ a complex constant
with $\eta=\Rs >0.$
\par
\medskip\noindent
{\it Proof.} If (4) is a solution, then
$$
\varphi_\gamma =e^{s\gamma t+i\gamma\om y} \varphi(\gamma x)
$$
is also a solution for any $\gamma>0.$ Thus we can find solutions which grow 
arbitrarily fast exponentially.

\medskip
We shall now discuss how to determine whether such solutions exist. We
introduce (4) into (1),(2) and obtain
$$ \eqalignno{
&\varphi_{xx}-(s^2+\om^2)\varphi=0,&(5)\cr
&s\varphi(0)=\alpha \varphi_x(0)+i\beta\om \varphi(0),\quad 
|\varphi|_\infty<\infty.&(6)\cr}
$$
(5),(6) is an eigenvalue problem. We can phrase Lemma 1 also as 
\proclaim
Lemma 1'. The problem is not well-posed if (5),(6) has an eigenvalue 
with $\Rs >0.$
\par
\noindent
(5) is an ordinary differential equation with constant coefficients and
its general solution is of the form
$$ \hat v=\sigma_1 e^{\kappa_1 x}+ \sigma_2 e^{\kappa_2 x},
\eqno(7) $$
where $\kappa_1=+\sqrt{s^2+\om^2},~
\kappa_2=-\sqrt{s^2+\om^2}$ are the solutions of the characteristic equation
$$ \kappa^2-(s^2+\om^2)=0. $$
We define the $\sqrt{}$ by
$$
-\pi < {\rm arg}(s^2+\om^2)\le \pi,\quad
 {\rm arg}\sqrt{s^2+\om^2}=\halv {\rm arg}(s^2+\om^2).
 \eqno(8) $$
Thus
$$ {\rm Re}\,\kappa_1>0\quad {\rm and}\quad 
{\rm Re}\,\kappa_2<0\quad {\rm for}\quad \Rs>0,~{\rm respectively}.
\eqno(9) $$
By assumption $\hat v$ is a bounded function and cannot contain
exponentially growing components. Therefore, by (9), this is only
possible if $\sigma_1=0,$ i.e.
$$ \varphi=\sigma_2 e^{\kappa_2 x}. \eqno(10) $$
Introducing (10) into the boundary condition gives us
$$ (s-\alpha\kappa_2-i\beta\om )\sigma_2=0. \eqno(11) $$
\smallskip\noindent
$\Rs >0$ and Re$\,\kappa_2<0$ tell us that there 
are no solutions for
$\Rs>0$ since, by assumption, $\alpha >0.$

For the purpose of proving well-posedness, from now
on we assume that the initial data (2) vanish. This
may always be achieved by the transformation 
$$
      u=v-e^{-t} f_1 - t e^{-t}(f_2+f_1). 
$$
(We start the time evolution from 'rest'.) Then we can solve (1)--(3) by
Fourier transform with respect to $y$ and Laplace transform with respect to $t$
and obtain the inhomogeneous versions of (5) and (6),
$$ \eqalign{
&\hat u_{xx}-(s^2+\om^2)\hat u =-\hat F,\quad \hat u=\hat u(x,\om,s),\cr
&(s-\alpha\kappa_2 -i\beta \om)\hat u(0,\om,s)=\hat g(\om,s).\cr}
\eqno(12)
$$
Since (5) and (6) have no eigenvalues for $\Rs>0,$ we can solve (12). By inverting
the Laplace and Fourier transform, we obtain a unique solution.

It is particularly simple to calculate  the solution for $\hat F\equiv 0.$
Corresponding to (10) and (11) we obtain
$$
\hat u=e^{\kappa_2 x}\hat u(0,\om,s) \eqno(13)
$$
where
$$
(s-\alpha\kappa_2 -i\beta \om)\hat u(0,\om,s)=\hat g(\om,s).
$$
To obtain sharp estimates we need two lemmas.

\proclaim
Lemma 2. There is a constant $\delta_1 >0$ such that
$$ \Reka={\rm Re}\,\sqrt{\om^2+s^2}\ge \delta_1 \eta,\quad s=i\xi+\eta,\quad \Rs=\eta.
 \eqno(14) $$
\par
\medskip\noindent
{\it Proof.} 
$$
\kappa=\sqrt{\om^2-\xi^2+2i\xi\eta +\eta^2}.
$$
Let
$$
\kappa'={\kappa\over\sqrt{\om^2+\xi^2}},\quad
\xi'={\xi\over\sqrt{\om^2+\xi^2}},\quad
\om'={\om\over\sqrt{\om^2+\xi^2}},\quad
\eta'={\eta\over\sqrt{\om^2+\xi^2}}.
$$
Then
$$
\kappa'=\sqrt{(\om')^2-(\xi')^2+2i\xi'\eta' +(\eta')^2}, \quad
(\om')^2+(\xi')^2=1.
$$
If $\eta'>\!> 1,$ then $\kappa'\approx\eta'$ and 
(14) holds. Thus we can assume that $(\om')^2+(\xi)^2+(\eta')^2 \le\con$
Assume that (14) is not true. Then there is a sequence 
$$
\om'\to \om'_0,\quad \xi'\to\xi'_0,\quad \eta'\to\eta'_0
$$
such that
$$ {\rm Re}\,{\kappa'\over \eta'}\to 0. \eqno(15) $$
This can only be true if $\eta'_0=0.$
 If $(\om'_0)^2>(\xi'_0)^2,$
then (15) cannot hold. If
  $(\om'_0)^2< (\xi'_0)^2,$ then
$(\xi')^2\ge\halv$ and, for sufficiently small $\eta',$
(8) gives
$$\eqalign{
\kappa'&\approx 
i\sqrt{(\xi_0')^2-(\om'_0)^2}+
{\xi_0'\over  \sqrt{(\xi_0')^2-(\om'_0)^2}} \eta',
\quad {\rm if}\quad \xi'_0>0,\cr
\kappa'&\approx 
-i\sqrt{(\xi_0')^2-(\om'_0)^2}-
{\xi_0'\over  \sqrt{(\xi_0')^2-(\om'_0)^2}} \eta',
\quad {\rm if}\quad \xi'_0<0,\cr}\eqno(16)
$$
and (14) holds. The same is true if $(\xi_0')^2=(\om_0')^2.$ This proves the
lemma.
\medskip\noindent
\proclaim
 Lemma 3. Assume that $\alpha >0$ and $|\beta| <1.$
There is a constant $\delta_2>0$ such that, for all $\om$ and $s$ with
$\Rs \ge 0,$
$$
|s-\alpha\kappa_2-i\beta\om|\ge\delta_2\sqrt{|s|^2+|\om|^2}.
\eqno(17) 
$$
\par
We use the same normalization as in Lemma 2 and write
(17) as
$$ \eqalign{
|L|&=:|i(\xi'-\beta\om')+\eta'+\alpha
\sqrt{(\om')^2-(\xi')^2+2i\xi'\eta'
+(\eta')^2} |\cr
&\ge \delta \sqrt{(\om')^2+(\xi')^2 +(\eta')^2} ,\quad
(\om')^2+(\xi')^2 =1. \cr}
\eqno(18)
$$
Since $\alpha >0$ and Re$\,\sqrt{s^2+\om^2}\ge 0,$ the inequality
holds for $\eta'>\!>1.$
Thus we can assume that $|\eta'|\le \con$ Assume now that there is no
$\delta>0$ such that (18) holds. Then there is a sequence
$\xi'\to \xi'_0,~\om'\to\om'_0,~\eta'\to 0$ such that
$$ L\to L_0 =0. \eqno(19) $$
Using (8) we obtain
$$
L\to L_0=\cases{
i(\xi'_0-\beta\om'_0)+\alpha 
\sqrt{(\om'_0)^2-(\xi'_0)^2}  & if $\om_0^2>\xi_0^2$\cr
&\cr
i(\xi_0-\beta\om_0)+
{\xi_0\over |\xi_0|}i\alpha\sqrt{(\xi'_0)^2-(\om'_0)^2} 
 & if $\xi_0^2\ge \om_0^2 .$\cr}
$$
Clearly, $L_0\ne 0$ if $|\beta|< 1.$ Thus we arrive at a contradiction
and (17) holds. This proves the lemma.

We can now prove
\proclaim
Theorem 1. There is a constant $K$ such that the solution (13) satisfies the estimates
$$ \eqalign{
|\hat u_x(0,\om,s)|&\le K|\hat g(\om,s)|,\cr
\sqrt{|s|^2+\om^2}\cdot
|\hat u(0,\om,s)|&\le K|\hat g(\om,s)|.\cr}
$$
Therefore we can use the theory of pseudo-differential operators to obtain
the estimates and results of the Appendix.
In particular, the problem is strongly well-posed in the generalized sense.
\par
\medskip\noindent
{\it Proof.} By (13) and (17),
$$
|\hat u_x(0,\om,s)|\le |\kappa_2|\,|\hat u(0,\om,s)|=
|\sqrt{\om^2+s^2}|\, |\hat u(0,\om,s)|\le K|\hat g(\om,s)|.
$$
The estimates for the other derivatives follow directly from (17) 
and (14). 

(A18) is the first order version of (12). For $F\equiv 0,$ we have $\hat u_x=
\sqrt{|s|^2+\om^2}\,\hat v$ and therefore, for $x=0,$
$$
\sqrt{|s|^2+|\om|^2}\, |\hat v(0,\om,s)|=|\hat u_x(0,\om,s)|\le K|\hat g(\om,s)|.
$$
Thus the required estimate (A20) holds and we can apply the Main theorems A1 and A2
and obtain the estimates (A21) and (A22).
This proves the theorem.

\def\part{\partial_t}
\def\parx{\partial_x}
\def\pary{\partial_y}
\def\parz{\partial_z}
\bigskip\noindent
\centerline{\bf 3. Linearized Einstein equations} 

\noindent
We consider the half-plane problem for the linearized Einstein equations
$$
\eqalign{ 
&(-\part^2+\parx^2+\pary^2+\parz^2)
\pmatrix{\gamma^{tt} & \gamma^{tx} & \gamma^{ty} & \gamma^{tz}\cr
\gamma^{tx} & \gamma^{xx} & \gamma^{xy} & \gamma^{xz}\cr
\gamma^{ty} & \gamma^{yx} & \gamma^{yy} & \gamma^{yz}\cr
\gamma^{tz} & \gamma^{zx} & \gamma^{zy} & \gamma^{zz}\cr} =F \cr
&\quad x\ge 0,~ t\ge 0,~ -\infty < y < \infty,~ -\infty < z < \infty,\cr}
\eqno(26)
$$
together with the constraints  $C^{\alpha}$,
$$ \eqalign{
C^t & =\part\gamma^{tt}+\parx\gamma^{tx}+\pary\gamma^{ty}+\parz\gamma^{tz}=0,\cr
C^x & =\part\gamma^{tx}+\parx\gamma^{xx}+\pary\gamma^{xy}+\parz\gamma^{xz}=0,\cr
C^y & =\part\gamma^{ty}+\parx\gamma^{yx}+\pary\gamma^{yy}+\parz\gamma^{yz}=0,\cr
C^z &
=\part\gamma^{tz}+\parx\gamma^{zx}+\pary\gamma^{zy}+\parz\gamma^{zz}=0.\cr}
\eqno(27)
$$
The constraints are also solutions of the wave equation. We can guarantee that
they remain zero at later times if $C^\alpha=0$, $\alpha=(t,x,y,z)$, are part
of the boundary conditions for (26) at $x=0.$

A possibility is
$$ \eqalign{
\part 
\pmatrix{
\gamma^{tt}\cr \gamma^{tx}\cr \gamma^{xx}\cr \gamma^{ty}\cr
\gamma^{xy}\cr \gamma^{tz}\cr \gamma^{xz}\cr
\gamma^{yy}\cr \gamma^{yz}\cr \gamma^{zz}\cr} &+\parx
\pmatrix{
0 & 1 & 0 & 0 & 0 & 0 & 0 & 0 & 0 & 0 \cr
0 & 0 & 1 & 0 & 0 & 0 & 0 & 0 & 0 & 0 \cr
a_1 & a_2 & a_3 & 0 & 0 & 0 & 0 & 0 & 0 & 0 \cr
0 & 0 & 0 & 0 & 1 & 0 & 0 & 0 & 0 & 0 \cr
0 & 0 & 0 & b_1 & b_2 & 0 & 0 & 0 & 0 & 0 \cr
0 & 0 & 0 & 0 & 0 & 0 & 1 & 0 & 0 & 0 \cr
0 & 0 & 0 & 0 & 0 & c_1 & c_2 & 0 & 0 & 0 \cr
0 & 0 & 0 & 0 & 0 & 0 & 0 & -1 & 0 & 0 \cr
0 & 0 & 0 & 0 & 0 & 0 & 0 & 0 & -1 & 0 \cr
0 & 0 & 0 & 0 & 0 & 0 & 0 & 0 & 0 & -1 \cr}
\pmatrix{
\gamma^{tt}\cr \gamma^{tx}\cr \gamma^{xx}\cr \gamma^{ty}\cr
\gamma^{xy}\cr \gamma^{tz}\cr \gamma^{xz}\cr 
\gamma^{yy}\cr \gamma^{yz}\cr \gamma^{zz}\cr} \cr
&+\pary
\pmatrix{
0 & 0 & 0 & 1 & 0 & 0 & 0 & 0 & 0 & 0 \cr
0 & 0 & 0 & 0 & 1 & 0 & 0 & 0 & 0 & 0 \cr
0 & 0 & 0 & 0 & 0 & 0 & 0 & 0 & 0 & 0 \cr
0 & 0 & 0 & 0 & 0 & 0 & 0 & 1 & 0 & 0 \cr
0 & 0 & 0 & 0 & 0 & 0 & 0 & 0 & 0 & 0 \cr
0 & 0 & 0 & 0 & 0 & 0 & 0 & 0 & 1 & 0 \cr
0 & 0 & 0 & 0 & 0 & 0 & 0 & 0 & 0 & 0 \cr
0 & 0 & 0 & 0 & 0 & 0 & 0 & 0 & 0 & 0 \cr
0 & 0 & 0 & 0 & 0 & 0 & 0 & 0 & 0 & 0 \cr
0 & 0 & 0 & 0 & 0 & 0 & 0 & 0 & 0 & 0 \cr}
\pmatrix{
\gamma^{tt}\cr \gamma^{tx}\cr \gamma^{xx}\cr \gamma^{ty}\cr
\gamma^{xy}\cr \gamma^{tz}\cr \gamma^{xz}\cr 
\gamma^{yy}\cr \gamma^{yz}\cr \gamma^{zz}\cr} \cr
&+\parz
\pmatrix{
0 & 0 & 0 & 0 & 0 & 1 & 0 & 0 & 0 & 0 \cr
0 & 0 & 0 & 0 & 0 & 0 & 1 & 0 & 0 & 0 \cr
0 & 0 & 0 & 0 & 0 & 0 & 0 & 0 & 0 & 0 \cr
0 & 0 & 0 & 0 & 0 & 0 & 0 & 0 & 1 & 0 \cr
0 & 0 & 0 & 0 & 0 & 0 & 0 & 0 & 0 & 0 \cr
0 & 0 & 0 & 0 & 0 & 0 & 0 & 0 & 0 & 1 \cr
0 & 0 & 0 & 0 & 0 & 0 & 0 & 0 & 0 & 0 \cr
0 & 0 & 0 & 0 & 0 & 0 & 0 & 0 & 0 & 0 \cr
0 & 0 & 0 & 0 & 0 & 0 & 0 & 0 & 0 & 0 \cr
0 & 0 & 0 & 0 & 0 & 0 & 0 & 0 & 0 & 0 \cr}
\pmatrix{
\gamma^{tt}\cr \gamma^{tx}\cr \gamma^{xx}\cr \gamma^{ty}\cr
\gamma^{xy}\cr \gamma^{tz}\cr \gamma^{xz}\cr 
\gamma^{yy}\cr \gamma^{yz}\cr \gamma^{zz}\cr} =g.\cr}
\eqno(28)$$
Here $a_1,~a_2,~a_3,~b_1,~b_2$ and $c_1,~c_2$ are real constants such that
the eigenvalues $\lambda_j$ of
$$
\pmatrix{ 0 & 1 & 0\cr 0 & 0 & 1 \cr a_1 & a_2 & a_3 \cr},\quad
\pmatrix{0 & 1\cr b_1 & b_2 \cr} \quad {\rm and}\quad
\pmatrix{ 0 & 1 \cr c_1 & c_2 \cr}
\eqno(29) $$
are real and negative.

We want to prove
\proclaim
Theorem 3. The half-plane problem for the system (26) with boundary
conditions (28) is well-posed in the generalized sense if the
eigenvalues of the matrix (29) are real and negative.
\par
\medskip\noindent
{\it Proof.}
We can assume that $F=0$ and need only show that the estimate of
Theorem 1 holds for every component $\gamma^{\mu\nu}$ with $|\hat g|$
denoting the Euclidean norm of all components of the forcing.

 Fourier transform the problem with respect to the 
tangential variables and Laplace transform it with respect to time.
For every component $\hat\gamma^{ij}$ we obtain
$$
\left(\parx^2-(s^2+\om_1^2+\om_2^2)\right)\hat\gamma^{\mu\nu}=
0,\quad |\hat\gamma|_\infty <\infty, \eqno(30) 
$$
which are coupled through the corresponding transformed boundary condition.
We start with the last component
$$ \eqalign{
&\left(\parx^2-(s^2+\om_1^2+\om_2^2)\right) \hat\gamma^{zz} =0,\quad x\ge 0,\cr
&s\hat\gamma^{zz}=\parx\hat\gamma^{zz}+\hat g^{zz}\quad {\rm for}\quad
x=0,\quad |\hat\gamma^{zz}|_\infty <\infty.\cr}
\eqno(31)
$$
This is a problem which we have treated in the last section.
By Theorem 1, we gain one derivative on the boundary.
For $\hat\gamma^{yz},~\hat\gamma^{yy}$ we have the same result.

The coupled boundary conditions for $\hat\gamma^{tz},~
\hat\gamma^{xz}$
$$ s\pmatrix{
\hat\gamma^{tz}\cr \hat\gamma^{xz}\cr}=
-\pmatrix{0 & 1 \cr c_1 & c_2 \cr} \parx
\pmatrix{
\hat\gamma^{tz}\cr \hat\gamma^{xz}\cr}+i\omega_1
\pmatrix{\hat\gamma^{yz}\cr 0\cr}+i\omega_2
\pmatrix{\hat\gamma^{zz}\cr 0\cr}+\pmatrix{\hat g^{tz}\cr \hat g^{xz}\cr} 
\eqno(32) $$
can be decoupled. There is a unitary matrix $U$ such that
$$ U^*\pmatrix{0 & 1 \cr c_1 & c_2 \cr}U=
\pmatrix{ -\lambda_1 & c_{12} \cr 0 & -\lambda_2 \cr},
\quad \lambda_1,\lambda_2 >0.
$$
Introducing new variables by
$$
\pmatrix{
\tilde\gamma^{tz}\cr \tilde\gamma^{xz}\cr} =
U \pmatrix{
\hat\gamma^{tz}\cr \hat\gamma^{xz}\cr}
$$
gives us
$$ s\pmatrix{ \tilde\gamma^{tz}\cr\tilde\gamma^{xz}\cr}=
\pmatrix{\lambda_1 & c_{12}\cr 0 & \lambda_2\cr}
\parx\pmatrix{ \tilde\gamma^{tz}\cr\tilde\gamma^{xz}\cr}+
U\left(i\omega_1\pmatrix{ \hat\gamma^{yz} \cr 0\cr}+
i\omega_2\pmatrix{ \hat\gamma^{zz} \cr 0\cr}+
\pmatrix{\hat g^{tz}\cr \hat g^{xz}\cr}\right) .
$$
We start with the last component. By (31), we have gained a derivative which we
loose by calculating $\parz \gamma^{zz}=i\omega_2 \gamma^{zz}.$ But then we
gain a derivative by solving for $\tilde\gamma^{xz}.$ The same is true for
$\tilde\gamma^{tz}.$ The process can be continued and the theorem follows.

Clearly, Theorem 3 is also valid when the matrices of (28) for the tangential
derivatives are strictly upper triangular, i.e., only terms above the diagonal
are not zero. This allows the sequential argument associated with (31) and (32)
above. One can also generalize the result to full matrices provided the
elements are sufficiently small.

\bigskip
\centerline{\bf 4. Constraint preserving boundary conditions in Sommerfeld form}
\medskip

The example in the preceding section illustrates how the pseudo-differential
theory can used to establish a constraint-preserving IBVP for the linearized
Einstein equations which is well-posed in the generalized sense. There are
further boundary conditions not covered in the example that can be treated by
the same technique because the boundary conditions can be written in the form
(28) with the matrices for the tangential derivatives strictly upper
triangular. Here we consider a simple hierarchy of Sommerfeld boundary
conditions. The geometric nature of the construction is more transparent using
standard tensor notation based upon spacetime coordinates
$x^{\alpha}=(t,x,y,z)$. Our results center about systems whose components
satisfy the scalar wave equation, now written in terms of the Minkowski metric
$\eta^{\alpha\beta}=diag(-1,1,1,1)$ as
$$ \eta^{\alpha\beta} \partial_\alpha \partial_\beta \Phi=
    ( -\partial_t^2 +\partial_x^2 + \partial_y^2+\partial_z^2)\Phi =0,
$$
in the half-space
$x\ge 0,~ t\ge 0,~ -\infty < y < \infty,~ -\infty < z < \infty$. 

For this IBVP, the energy 
$$
   E ={1\over 2}\int \bigg( (\partial_t \Phi)^2 +(\partial_x \Phi)^2 
    +(\partial_y \Phi)^2 +(\partial_z \Phi)^2 \bigg ) dx dy dz
$$
satisfies
$$
   \partial_t E = -\int_{x=0} {\cal F} dy dz
$$
where the energy flux through the boundary is
$$
  {\cal F} = (\partial_t \Phi) \partial_x \Phi. 
$$
This leads to a range of homogeneous, dissipative boundary conditions 
$$
    A \partial_t + B  \partial_x \Phi =0,
$$
subject to values of $A$ and $B$ such that ${\cal F} \ge 0$. The Dirichlet
boundary condition $(A=1,B=0)$ corresponds to the case where $\Phi$ has an odd
parity local reflection symmetry across the boundary; and the Neumann boundary
condition $(A=0,B=1)$ corresponds to an even parity local reflection symmetry.
Both the homogeneous Dirichlet and Neumann conditions are borderline
dissipative cases for which ${\cal F}=0$. The description of a traveling wave
carrying energy across the boundary requires an inhomogeneous form of the
Dirichlet or Neumann boundary condition, which provides the proper boundary data
for the wave to pass through the boundary. However, in numerical simulations,
such inhomogeneous Dirichlet or Neumann boundary data can only be prescribed
for the signal and the numerical error is reflected by the boundary and
accumulates in the grid. This can lead to poor performance in simulations of
dynamical nonlinear systems such as Einstein's equations. For such
computational purposes, it is more advantageous to use the strictly dissipative
Sommerfeld condition given by $(A=1,B=-1)$, i.e. 
$$
     (\partial_t -\partial_x)\Phi =0,
$$
for which ${\cal F}=(\partial_t \Phi)^2$. The Sommerfeld condition is based
upon the characteristic direction in the 2-space picked out by the outward
normal to the boundary and a timelike direction (the evolution direction)
tangent to the boundary.

Before formulating Sommerfeld boundary conditions for the constrained
harmonic Einstein equations, it is instructive to consider the
analogous case of Maxwell's equations for the electromagnetic field
expressed in terms of a vector potential $A^\mu=(A^t,A^x,A^y,A^z)$ (see
e.g. [17]). Subject to the Lorentz gauge condition
$$
   C:= \partial_\mu A^\mu =0 ,
\eqno(33)
$$
the Maxwell equations reduce to the wave equations
$$
     \eta^{\alpha\beta} \partial_\alpha \partial_\beta A^\mu =0 .
\eqno(34)
$$
The constraint $C$ also satisfies the wave equation $\eta^{\alpha\beta}
\partial_\alpha \partial_\beta C=0$. Thus Cauchy data $A^\mu|_{t=0}$ and 
$\partial_t A^\mu |_{t=0}$ which satisfies ${\cal C}|_{t=0}=\partial_t{\cal
C}|_{t=0} =0$, i.e. for which $C$ also has vanishing Cauchy data, leads to the
well-known result that the constraint is preserved for the Cauchy problem. 

In order to extend constraint preservation to the IBVP, the boundary condition
for $A^{\mu}$ must imply a homogeneous boundary condition for ${\cal C}$. One
way to accomplish this is to use locally reflection symmetric boundary data,
as in the scalar case. For instance, for the above half-space problem, the
even parity boundary conditions $\partial_x A^t|_{x=0}= \partial_x
A^y|_{x=0}=\partial_x A^z|_{x=0}=A^x|_{x=0} =0 $ imply that $\partial_x
C|_{x=0} =0 $. These boundary conditions are dissipative and lead to a
well-posed IBVP for Maxwell's equations. The analogous approach has been used
to formulate a well-posed constraint-preserving IBVP for the nonlinear
Einstein equations [13]. However, this leads to a combination of Dirichlet and
Neumann boundary conditions which is only borderline dissipative and the
results of tests for the nonlinear Einstein problem show that a strictly
dissipative Sommerfeld-type boundary condition gives better numerical accuracy
[15].

The pseudo-differential theory offers alternative approaches.
Consider the IBVP with Sommerfeld boundary conditions 
$$
     (\partial_t -\partial_x)( A^t+A^x)=g, 
\eqno(35)
$$
$$
     (\partial_t -\partial_x)  A^y = g^y,  
\eqno(36)
$$
$$
      (\partial_t -\partial_x)  A^z =g^z.  
\eqno(37)
$$
$$
      {1\over 2} (\partial_t -\partial_x)(A^t-A^x)+
        \partial_t(A^t+A^x) +\partial_y A^y
       +\partial_z A^z ={1\over 2}g,
\eqno(38)
$$
where $g$, $g^y$ and $g^z$ are free Sommerfeld data. The IBVP is well-posed in
the generalized sense if the estimates of Theorem 1 are satisfied.  Using the
argument associated with (31) in the preceding section, it follows from
(35)--(37) that $\hat A^t+\hat A^x$, $\hat A^y$, $\hat A^z$ and their
derivatives satisfy these estimates.  Next, consider the Sommerfeld boundary
condition (38) for $A^t-A^x$. Using the argument associated with (32), the
tangential derivatives of $(A^t+A^x)$, $A^y$ and $A^z$ introduce no problem,
and the estimates extend to  $\hat A^t-\hat A^x$ and its derivative. Thus the
Sommerfeld boundary conditions (35)--(38) guarantee a well-posed IBVP for the
system (34). But (35)--(38) also imply that the constraint satisfies the
homogeneous boundary condition $C=0$, as is evident by rewriting (33) in the
form
$$
   C= {1\over 2} (\partial_t -\partial_x)(A^t-A^x)+
       {1\over 2} (\partial_t +\partial_x)(A^t+A^x) +\partial_y A^y
       +\partial_z A^z.
\eqno(39)
$$

Although the boundary conditions (35)--(38) lead to a well-posed,
constraint preserving IBVP for Maxwell's equations, they do not
correspond to any physically familiar boundary conditions on the electric and
magnetic field vectors ${\bf E}$ and ${\bf B}$. However, there are numerous
options in constructing boundary conditions by this approach. For instance,
consider the choice
$$
     (\partial_t -\partial_x)( A^t+A^x)=0,  
\eqno(40)
$$
$$      
     (\partial_t -\partial_x)  A^y + \partial_y (A^t+A^x) = 0, 
\eqno(41)
$$
$$
      (\partial_t -\partial_x)  A^z + \partial_z (A^t+A^x) = 0.  
\eqno(42)
$$
$$
      {1\over 2} (\partial_t -\partial_x)(A^t-A^x)+
       \partial_t (A^t+A^x) +\partial_y A^y
       +\partial_z A^z = 0.
\eqno(43)
$$
Again, the Sommerfeld condition (40) implies that ${\hat A}^t+{\hat A}^x=0$ and
its derivatives satisfy the estimates required for Theorem 1. In addition, the
sequential manner in which the tangential derivatives of previously
estimates quantities enter (41)--(43) ensures that all components of ${\hat
A}^\mu$ satisfy the required estimates. Furthermore, (40)--(43) imply that the
constraint satisfies the homogeneous boundary condition $C=0$, as is again
evident from (39). By using the wave equations (34) and the constraint (33), it
can be verified that these boundary conditions give rise to the familiar plane
wave relations $E^y=-B^z$ and $E^z=B^y$ on the components of the electric and
magnetic fields. These relations imply that the Poynting flux ${\bf E}\times
{\bf B}$ leads to a loss of electromagnetic energy from the system.

The half-space problem for the linearized
harmonic Einstein equations can be treated in a similar manner.
Given a metric tensor $g_{\mu\nu}$ with inverse $g^{\mu\nu}$
and determinant $g$, the linearized Einstein
equations imply that the perturbations
$\gamma^{\mu\nu}=\delta (\sqrt{-g} g^{\mu\nu})$ satisfy the 10 wave equations
$$
   \eta^{\alpha\beta} \partial_\alpha \partial_\beta  \gamma^{\mu\nu} =0 
 \eqno(44)
$$
 provided the 4 constraints 
$$
   {\cal C}^\mu:= \partial_\nu \gamma^{\mu\nu}=0
  \eqno(45)
$$
are satisfied [17]. The constraints constitute the harmonic gauge conditions
which reduce the Einstein equations to a symmetric hyperbolic system
of wave equations.

A simple formulation of Sommerfeld boundary conditions for
the half-space problem 
$0\le x<\infty$  for the linearized gravitational field
can be patterned after the above treatment of the electromagnetic
problem. For convenience we write $x^A=(y,z)$.
First we require the 6 Sommerfeld boundary conditions
$$
  (\partial_t -\partial_x)\ \gamma^{AB}=q^{AB},
\eqno(46)
$$
$$
  (\partial_t -\partial_x)( \gamma^{tA}+\gamma^{xA} )=q^A,
\eqno(47)
$$
$$
  (\partial_t -\partial_x)( \gamma^{tt}+2\gamma^{tx}+\gamma^{xx} )=q,
\eqno(48)
$$ 
where $q^{AB}$, $q^A$ and $q$ are free Sommerfeld data.
Then the constraints are used to supply 4 additional Sommerfeld-type
boundary conditions in the hierarchical order
$$
  {\cal C}^A={1\over 2}(\partial_t -\partial_x)(\gamma^{tA}-\gamma^{xA})
      +\partial_t(\gamma^{tA}+\gamma^{xA})+\partial_B\gamma^{AB} 
      -{1\over 2}q^{AB}=0,
\eqno(49)
$$
$$
  {\cal C}^t+ {\cal C}^x= {1\over 2} (\partial_t -\partial_x)
                         (\gamma^{tt}-\gamma^{xx}) 
	      +\partial_t( \gamma^{tt}+2\gamma^{tx}+\gamma^{xx} )
	      +\partial_B(\gamma^{tB} + \gamma^{xB} ) -{1\over 2}q =0,
\eqno(50)
$$
$$
     {\cal C}^t= {1\over 2} (\partial_t-\partial_x)
            (\gamma^{tt}+\gamma^{xx})
               +\partial_t( \gamma^{tt}+\gamma^{tx} )
	      +\partial_B\gamma^{tB}-{1\over 2}q =0 .
\eqno(51)
$$ 
In expressing the constraints in this form, we have used (47)--(49) and the
prior constraints in the hierarchy.

The sequence of Sommerfeld conditions (46)--(51) for
$(\partial_t-\partial_x) \gamma^{\mu\nu}$ have the property that all
tangential derivatives of $\gamma^{\mu\nu}$ only involve prior
components in the sequence. This allows us to again use the arguments
associated with (31) and (32) to obtain the estimates for
all components of  $\hat \gamma^{\mu\nu}$ and their derivatives which
are required for Theorem 1.

The boundary conditions (46)--(51) offer a simple and attractive scheme for
numerical use. The pseudo-differential theory allows many more possibilities
for a well-posed, constraint-preserving IBVP. It would be of value to find a
version with a simple physical interpretation in the analogous way  that
(40)--(43) implies a positive Poynting flux in the electromagnetic case.
However, this issue is complicated by the lack of a unique expression for the
gravitational energy flux except in the asymptotic limit of null infinity. One
practical alternative for a numerical scheme would be to use an external
solution to provide the Sommerfeld data on the right hand sides of
(46)--(48). This data can be obtained either by matching to
an external linearized solution or, in the nonlinear case, by
Cauchy-characteristic matching [18].

Our results generalize to the curved space linearize harmonic wave equation,
whose principle part has the form
$$
      g^{\alpha\beta}\partial_\alpha\partial_\beta \gamma^{\mu\nu},
\eqno(52)
$$   
determined by a given space-time metric  $g^{\alpha\beta}(t,x,y,z)$, i.e. a
matrix which can be transformed at any point to diagonal Minkowski form.  This
variable coefficient problem is well-posed in the generalized sense if all
frozen coefficient problems are well-posed [5]. This result is insensitive to
lower order terms in the equations. This {\it principle of frozen coefficients}
is an important result of the pseudo-differential theory. We can reduce the
frozen coefficient problem to the above Minkowski space problem by adapting the
harmonic coordinates so that the boundary is given by $x=0$ and then
introducing a orthonormal tetrad,
$$
     g_{\alpha\beta}=-T_{\alpha} T_{\beta} + X_\alpha X_\beta
     +Y_\alpha Y_\beta +Z_\alpha Z_\beta,
$$
where $X_\alpha$ is in the direction $\nabla_\alpha x$ normal to the boundary.
Freezing the tetrad at a boundary point, we can then introduce
the linear coordinate transformation
$$
    \tilde t = T_{\alpha}x^{\alpha}, \,  \tilde x = X_{\alpha}x^{\alpha},
    \,  \tilde y = Y_{\alpha}x^{\alpha},  \,  \tilde z = Z_{\alpha}x^{\alpha}.      
$$
In the $\tilde x^\alpha$ coordinates, the frozen coefficient problem
reduces to the Minkowski space problem, which we have treated.

The full treatment of the Einstein equations requires taking into account the
relation $\gamma^{\mu\nu}=\sqrt{-g}g^{\mu\nu}$, which converts (52) into a
quasi-linear operator. In this case the pseudo-differential theory outlined in
the Appendix establishes that the IBVP is well-posed locally in time.

\bigskip\noindent
\centerline{\bf  Appendix}
\medskip
In this section we shall give a short summary of the theory for first order
systems. This includes also second order systems because our theory is based on
pseudo-differential operators and therefore one can always write a second order
system of differential equations in terms of a first order system of
pseudo-differential operators. (See the references in the Introduction.)

Consider a first order system
$$
u_t=P(\ipartx)u+F,\quad P(\ipartx)=A\, \ipartx_1+\sum_{j=2}^m B_j \ipartx_j
\eqno(A1) $$
with constant coefficients on the half-space
$$ t\ge 0,\quad x_1\ge 0,\quad -\infty < x_j <\infty, ~j=2,\ldots, m.
$$
Here $u(x,t)=\left(u^{(1)}(x,t),\ldots,u^{(n)}(x,t)\right)$ is a
vector valued function of the real variables $(x,t)=(x_1,\ldots, x_m,t)$ 
and $A,B_j$ are constant $n\times n$ matrices.

We assume that the system is strictly hyperbolic, i.e., for all real $\om=(\om_1,
\om_-),~\om_-=(\om_2,\ldots,\om_m)$ with $|\om|=1,$ the eigenvalues of the symbol
$$ P(i\om)=iA\om_1+iB(\om_-),\quad B(\om_-)=\sum_{j=2}^m B_j \om_j ,
\eqno(A2) $$
are purely imaginary and distinct.

In [1] the theory has been extended to the case where the eigenvalues  have
constant multiplicity and there is a complete set of eigenvectors. In particular,
the theory holds if the system consists of strictly hyperbolic subsystems
which are only coupled through lower order terms and the boundary conditions.

We assume also that $A$ is nonsingular and without restriction we can assume
that it has the form
$$ A=\pmatrix{-\Lambda^I & 0\cr 0& \Lambda^{II}\cr}.\eqno(A3) $$
Here $\Lambda^I,\Lambda^{II}$ are real positive definite diagonal matrices
of order $r$ and $n-r,$ respectively. For the singular case, see [7].

For $t=0,$ we give initial data
$$ u(x,0)= f(x)  \eqno(A4) $$
and for, $x_1=0,$ $r$ boundary conditions
$$
u^I(0,x_{-},t)=Su^{II}(0,x_{-},t)+g(x_{-},t),\quad x_{-}=(x_2,\ldots,x_m).
\eqno(A5) $$
All data are smooth, compatible and have compact support.

The usual theory for well-posed problems depends strongly on the assumption
that the system is symmetric and that the boundary conditions are maximally
dissipative. If any of those two arguments is not satisfied, then the theory
 does not give anything. We want to discuss a concept of well-posedness for which
we obtain necessary and sufficient conditions.

The main ingredient of a definition for well-posedness is the estimate of the 
solution in terms of the data. (See [5,Sec.7.3]). We will consider (A1),(A4),(A5)
with homogeneous initial data $f\equiv 0$ and use
\proclaim
Definition A1. Let $f(x)\equiv 0.$ We call the problem strongly well-posed in the
generalized sense if, for all smooth compatible data $F,g,$ there is a unique
solution $u$ and in every time interval $0\le t\le T$ there is a constant $K_T$
which does not depend on $F$ and $g$ such that
$$
\int_0^t \|u(\cdot,\tau)\|^2 d\tau + 
\int_0^t \|u(\cdot,\tau)\|^2_- d\tau  \le
K_T \{ \int_0^t \|F(\cdot,\tau)\|^2 d\tau +
\int_0^t \|g(\cdot,\tau)\|^2_- d\tau \}. 
\eqno(A6)
$$
Here $\|\cdot\|,\|\cdot\|_-$ denote the $L_2$-norm with respect to the half-space
and the boundary space, respectively.

We start with a simple observation. For $F=g=0,$ we construct simple wave
 solutions
$$u(x_1,x_-,t)=e^{st+i\langle \om,x\rangle_-} \varphi(x_1),\quad
\langle \om,x\rangle_-=\sum_{j=2}^m \om_j x_j, \eqno(A7)
$$
satisfying the boundary conditions
$$ \varphi^I(0,x_-)=S\varphi^{II}(0,x_-),\quad |\varphi|_\infty <\infty.
\eqno(A8) $$
We have
\proclaim
Lemma A1. The half-plane problem is not well-posed if, for some $\om_0$
and complex $s_0$ with $\Rs_0>0,$ 
there is a solution (A7) which satisfies (A8).
\par
\medskip\noindent
{\it Proof.} If there is a solution then, by homogeneity,
$$
u(\gamma x_1,\gamma x_-,\gamma t)=e^{\gamma (s_0 t+i\langle \om_0,x \rangle_- )}
 \varphi(\gamma x_1),
$$
is also a solution for any $\gamma >0.$ Thus there are solutions which grow 
arbitrarily fast exponentially.
This proves the lemma.
\medskip
We shall now derive algebraic conditions so that we can decide whether these
simple wave solutions exist.

Introducing (A7) into (A1) and (A5), we obtain
$$\eqalign{
 s\varphi&=A\varphi_x+iB(\om_-)\varphi,\quad x\ge 0, \cr
\varphi^I(0)&=S\varphi^{II}(0),\quad |\varphi|_\infty <\infty. \cr}
\eqno(A9) $$
(A9) is an eigenvalue problem for a system of ordinary differential equations
which can be solved in the usual way. Let $\kappa$ denote
the solutions of the characteristic equation
$$ {\rm Det}|A\kappa-\left(sI-iB(\om_-)\right)|=0. \eqno(A10) $$
One can prove
\proclaim
Lemma A2.
\smallskip\noindent
1) For $\Rs >0,$ there are no $\kappa$ with  $\Reka =0.$
\smallskip\noindent
2) There are exactly $r$ eigenvalues with $\Reka <0$ and $n-r$ eigenvalues with
$\Reka>0.$
\smallskip\noindent
3) There is a constant $\delta>0$ such that, for all $s=i\xi+\eta,~\xi,\eta >0,$
and all $\om_-,$
$$ |\Reka |>\delta\eta,\quad  s=i\xi+\eta,~ \xi,\eta >0~{\rm real.} $$
 (See [2]).
\par

We can now write down the general solution of (A9). 
If all $\kappa_j$ are distinct, the solution is of the form
$$ \varphi=\sum_{\Reka_j<0} \sigma_j e^{\kappa_jx} h_j+
\sum_{\Reka_j>0} \sigma_j e^{\kappa_jx} h_j. \eqno(A11)
$$
Here $h_j$ are the corresponding eigenvectors.
(If the eigenvalues are not distinct, the usual modifications apply.)

Since we are only interested in bounded solutions, all $\sigma_j$ in the second
term are zero. Introducing $\varphi$ into the boundary conditions at $x=0$ gives
us a linear system of $r$ equations for $r$ unknowns $(\sigma_1,\ldots,\sigma_r)=
{\underline{\sigma}}$ which we write as
$$ C(\om_-,s) {\underline{\sigma}} =0. \eqno(A12) $$

The problem is not well-posed if for some $\om_-$, there is an eigenvalue $s_0$
with $\Rs_0>0,$ i.e., Det$\,C(\om_-,s_0)=0.$ Then the linear system of
equations (A12) and therefore also (A9) has a nontrivial solution.

From now on we shall assume that Det$\,C\ne 0$ for $\Rs >0.$ Then we can solve
the initial boundary value problem by Laplace transform in time and Fourier
transform in the tangential variables. For convenience, we start the solution
from 'rest', i.e., $u(x,0)=f(x)\equiv 0.$ Then we obtain
$$ \eqalign{
s\hat u&=A\hat u_x+iB(\om_-) \hat u+\hat F,\cr
\hat u^I(0)&=S\hat u^{II}(0)+\hat g.\cr} \eqno(A13)
$$
Since, by assumption, (A9) has only the trivial solution for $\Rs>0$ and
$|\Reka|>\delta\eta$, (A12) has a unique solution. Inverting the Fourier and
Laplace transforms we obtain the solution in physical space.

It is particularly simple to solve (A13) if $F\equiv 0.$ 
By (A11) and (A12),
$$ \hat u(0,\om_-,s)=\sum_{\Reka_j <0} \sigma_j e^{\kappa_j  x} h_j, $$
where the $\sigma_j$ are determined by
$$ C(\om_-,s) {\underline{\sigma}}=\hat g. $$
\proclaim
Definition A2.
Consider (A13) with $\hat F\equiv 0.$
 The problem is called boundary stable if, for all 
$\om,s$ with $\eta=\Rs>0,$ there is a constant $K$ which does not depend
on $\om,s$ and $\hat g$ such that
$$ |\hat u(0,\om,s)|\le K|\hat g(\om,s)|.  \eqno(A14) $$
\par
\medskip
One can also phrase the condition as: The eigenvalue problem (A9) has
no eigenvalues for $\Rs\ge 0$ or Det$\,C(\om_-,s)\ne 0$ for
$\Rs\ge 0.$

The estimate (A14) is crucial to the theory. It allows us to construct
 a symmetrizer to obtain an energy estimate in the generalized sense
for the full problem.

We introduce normalized  variables
$$
s'=s/\sqrt{|s|^2+|\om_-|^2}\,=i\xi'+\eta',\quad
\om_-'=\om_-/\sqrt{|s|^2+|\om_-|^2} $$
and write (A13) as
$$
\eqalign{
-A\hat u_x+ \sqrt{|s|^2+|\om_-|^2} 
\left(s' I-iB(\om'_-)\right)\hat u &=\hat F,\cr
\hat u^I(0)-S\hat u^{II}(0)&=\hat g.\cr} \eqno(A15)
$$
\proclaim
Main theorem A1. Assume that the half-plane problem is boundary stable. Then
there exists a symmetrizer $\hat R=\hat R(s',\om'_-)$ with the
following properties.
\smallskip\noindent
1) $\hat R$ is a smooth bounded function of $s',\om'_-$ and the
coefficients of $A,B_j$ and $S.$
\smallskip\noindent
2) $\hat R A$ is Hermitian and for all vectors $y$ which satisfy the
boundary conditions
$$ \langle y,\hat RAy\rangle \ge \delta_1|y|^2-c|g|^2.$$
\smallskip\noindent
3) $\sqrt{|s|^2+|\om_-|^2}\,
 {\rm Re}\, \{\hat R
\left(s' I-iB(\om'_-)\right)\}\ge \delta_2\eta I.$

\noindent
Here $\delta_1,\delta_2>0,~c>0$ are constants independent of 
$s',\om'_-.$
\par

We can now prove
\proclaim
Main theorem A2. Assume that the half-plane problem is boundary stable. 
Then it is well-posed in the sense of Definition A1.
\par
\medskip\noindent
{\it Proof.}
 Multiplying (A15) by 
$\hat R$ we obtain, for the $L_2$ scalar product with respect
to $x_1,$
$$ \eqalign{
{\rm Re}(\hat u,\hat R\hat F)&=
{\rm Re}\{-(\hat u,\hat R A\partial \hat u/\partial x_1)+
\left(\hat u,
\sqrt{|s|^2+|\om_-|^2}\cdot\hat R(s' I-iB(\om'_-)\hat u\right)\}\cr
&=
{\rm Re}\{-{1\over 2}\langle \hat u,\hat R A\hat u\rangle \big|_{x_1=0}^\infty
+\left(\hat u,
\hat R(sI-iB(\om_-)\hat u\right)\}\cr
&\ge \delta_1|\hat u(0,s,\om_-)|^2+\delta_2\eta \|\hat u(x_1,\om_-)\|^2-
c|\hat g|^2.\cr}
$$
Thus we obtain
$$
\eta \|\hat u(x_1,s,\om)\|^2+|\hat u(0,s,\om)|^2\le
\con\left( {1\over\eta} \|\hat F\|^2+c|\hat g|^2\right).
\eqno(A16) 
$$
Inverting the Laplace and Fourier transform proves the theorem.
\medskip\noindent
{\bf Remark}. The estimate (A14) and the properties of the symmetrizer and
therefore also the estimate (A16) need only be valid for $\eta\ge \eta_0 >0.$
This is important if lower order terms or variable coefficients are present.
\medskip
One could have derived the estimate (A16) by directly calculating
 the solution of (A9). However, the importance of the symmetrizer is that
we can consider $\hat R$ as a symbol and (A9) as an equation for
pseudo-differential operators. The theory of pseudo-differential operators
has far reaching consequences. In particular, the computational rules
for pseudo-differential operators show:
\smallskip\noindent
1) The estimate is also valid if the symbols depend smoothly on $x,t,$
provided we assume that $\eta>\eta_0,~\eta_0 $ sufficiently large.
Here $\eta_0$ depends on a finite number of $x,t$-derivatives of the symbols.
Therefore we extend the estimate to systems with variable coefficients.
\smallskip\noindent
2)  Well-posedness will not be destroyed by lower order terms. Therefore
one can localize the problem, and well-posedness in general domains can be
reduced to the study of the Cauchy problem and the half-plane problems.
\smallskip\noindent
3) The principle of frozen coefficients holds. 
\smallskip\noindent
4) The properties of the pseudo-differential operators allow us to estimate
derivatives in the same way as for standard partial differential equations.
Therefore we obtain well-posedness in the generalized  sense  for linear systems
with variable coefficients which gives us the corresponding local results for
quasi-linear problems.

\medskip
Since pseudo-differential operators are much more flexible than standard
differential operators, we can always write second order systems as first
order systems. Consider, for example, the problem we discuss in Section 2.
$$\eqalign{
&u_{tt}=u_{xx}+u_{yy}+F\quad {\rm for}\quad x\ge 0,~-\infty <y<\infty,~ t\ge 0,\cr
&u_t-\alpha u_x-\beta u_y=g,\quad x=0,~-\infty < y<\infty,~t\ge 0.\cr}
\eqno(A17)
$$
After Laplace-Fourier transform it becomes
$$\eqalign{
&\hat u_{xx}=(s^2+\om^2)\hat u-\hat F,\cr
& s\hat u-\alpha \hat u_x-\beta i\om \hat u =\hat g.\cr}
$$
Introducing a new variable $\hat u_x=\sqrt{|s|^2+\om^2}\,\hat v$
gives us the first order system
$$\eqalign{
& \hat{\bbf u}_x=
\sqrt{|s|^2+\om^2}\,
\pmatrix{ 0 & 1\cr s^{'2}+\om^{'2} & 0 \cr}
 \hat{\bbf u} -\tilde F,\quad 
 \hat{\bbf u}=\pmatrix{\hat u\cr \hat v\cr}, \cr
&\cr
&s'\hat u-\alpha \hat v-\beta i\om'\hat u=\tilde g,\cr}
\eqno(A18)
$$
where
$$
\tilde F={1\over\sqrt{|s|^2+\om^2}} \pmatrix{0\cr \hat F\cr},\quad
\tilde g={1\over\sqrt{|s|^2+\om^2}} \hat g.
\eqno(A19)
$$
\proclaim
Definition A3. We call the problem (A17) strongly well-posed
in the generalized sense if the corresponding first order problem (A18) with
general data $\tilde F,\tilde g$ has this property.
\par
\medskip
If (A18) is boundary stable, i.e., if the estimate (A14) holds with
$\hat g$ replaced by $\tilde g,$ then we can use the same technique as in
[2, Sec.4] to construct the symmetrizer. Therefore Main Theorems 1 and 2 are
valid and we obtain the estimate (A16) with $\hat F,\hat g$ replaced by
$\tilde F,\tilde g.$

Starting from (A17), $\tilde F,\tilde g$ satisfy (A19) and therefore (A14) becomes
$$
|\hat u(0,s,\om)|+|\hat v(0,s,\om)|\le {K\over \sqrt{|s|^2+\om^2}}
\, |\hat g(\om,s)|. \eqno(A20)
$$
In Section 2 we prove that (A20) holds by directly calculating the solution
of (A17) for $\hat F= 0.$

Similarly, in terms of $\hat F$ and $\hat g$ for the second order system, the
estimate (A16) becomes
$$
(|s|^2+\om^2)\,(\eta \|\hat{\bbf u}(\cdot,\om,s)\|^2+
|\hat{\bbf u}(0,s,\om)|^2)\le\con \left({1\over\eta} \|\hat F\|^2+
c|\hat g|^2 \right). \eqno(A21)
$$
We have also
$$
\|\hat u_x\|^2=(|s|^2+\om^2)\,\|\hat v\|^2 \le (|s|^2+\om^2)\|\hat{\bbf u}\|^2.
 \eqno(A22) $$
Therefore, by inverting the Laplace and Fourier transform, we can estimate the
$L_2$-norm of all first derivatives in terms of the $L_2$-norm of the data.
Thus we gain one derivative.

\bigskip
\centerline{\bf Acknowledgments}
\medskip
We thank H. Friedrich and O. Reula for many valuable discussions. This work was
partially supported by the National Science Foundation under grant PH-0244673 to
the University of Pittsburgh.

\bigskip
\centerline{\bf References}
\medskip\noindent
\item{[1]} Agranovich, M.S. (1972). "Theorem of matrices depending on parameters
and its application to hyperbolic systems", {\it Functional Anal. Appl.}
{\bf 6}. pp. 85-93.
\smallskip\noindent
\item{[2]} Kreiss, H.-O. (1970), "Initial boundary value problems for
hyperbolic systems". 
\item{} {\it Comm. Pure Appl. Math.} {\bf 23}. pp. 277-298.
\smallskip\noindent
\item{[3}] Rauch, J. (1972)," $L_2$ is a continuable condition for Kreiss'
mixed problems". {\it Comm. Pure Appl. Math.} {\bf 25}. pp. 265-285.
\smallskip\noindent
\item{[4}] Rauch, J. (1972),"Energy and resolvent inequalities for hyperbolic mixed
systems". {\it J. Differential Equations} {\bf 11}. pp. 528-450.
\smallskip\noindent
\item{[5]} Kreiss, H.-O. and Lorenz, J., "Initial-Boundary Value Problems and the
Navier-Stokes Equations", 
reprinted as {\it SIAM Classic}, 2004.
\smallskip\noindent
\item{[6]} Gustafsson, B., Kreiss, H.-O. and Oliger, J. "Time Dependent Problems and
Difference Methods",  1995 {\it Wiley \& Sons}
\smallskip\noindent
\item{[7]} Majda, A. and Osher, S. (1984), ``Initial-boundary value problems for
hyperbolic equations with uniformly characteristic boundary´´, 
{\it Comm. Pure Appl. Math.} {\bf 28}, pp. 607-675.
\smallskip\noindent
\item{[8]} Kreiss, H.-O. and Ortiz, O.E., (2002), {\it Lect. Notes Phys.} {\bf 604}, 359.
\smallskip\noindent
\item{[9]} Foures-Bruhat, Y. (1952),
"Theoreme d'existence pour certain systemes d'equations aux
derive\`es partielles nonlinaires",
{\it Acta Math} {\bf 88}, 141.
\smallskip\noindent
\item{[10]} Stewart, J.M. (1998), "The Cauchy problem and the initial boundary
value problem in numerical relativity", {\it Class. Quantum Grav.} {\bf 15},
2865.
\smallskip\noindent
\item{[11]} Reula, O. and Sarbach, O. (2005), "A model problem for the initial
boundary value formulation of Einstein's field equations", {\it J. of
Hyperbolic Differential Equations} {\bf 2}, 397. 
\smallskip\noindent
\item{[12]}  Sarbach, O. and Tiglio, M. (2005), "Boundary conditions for
Einstein's equations: Mathematical and numerical analysis", {\it J. of
Hyperbolic Differential Equations} {\bf 2}, 839. 
\smallskip\noindent
\item{[13]} Szil\`agyi, B. and Winicour, J.  (2003),
``Well-posed initial-boundary evolution in general relativity'', 
{\it Phys. Rev.} {\bf D68}, 041501.
\smallskip\noindent
\item{[14]} Babiuc, M.C., Szil\`agyi, B. and Winicour, J. (2006), ``Some
mathematical problems in numerical relativity'', {\it Lect. Notes Phys.} {\bf
692}, 251-274.
\smallskip\noindent
\item{[15]}Babiuc, M.C., Szil\`agyi, B. and Winicour, J., ``Testing numerical
relativity with the shifted gauge wave'', gr-qc/0511154.
\smallskip\noindent
\item{[16]} Friedrich, H. and Nagy, G. (1999), {\it Commun. Math. Phys.}, {\bf 201}, 619.
\smallskip\noindent
\item{[17]} Wald, R.~M. "General Relativity",  1984
{\it University of Chicago Press}. 
\smallskip\noindent
\item{[18]} Winicour, J. (2005), ``Characteristic evolution and matching'',
{\it Living Rev. Relativity} {\bf 8}, 10.

\bye

%% file: macros.tex
\magnification=1200
\baselineskip=16pt
\nopagenumbers
\headline={\hss\tenrm -- \sec\folio --\hss}

\def\halv{{1\over 2}}
\def\p2{2\pi}

\def\gp2{{1\over \p2}}

\def\ipartx{\partial /\partial x}
\def\partt{{\partial \over\partial t}}

\def\om{\omega}

\def\bbf#1{{\bf #1}}

\def\con{\;{\rm const.}\;}
\def\'{^\prime}
\def\Rs{\;{\rm Re}\;s}

\def\snok2#1{\hbox{$\tilde{\mkern -2.0mu \tilde #1} $}}

\def\Reka{{\rm Re}\,\kappa}